\documentclass[11pt]{article}

\usepackage[preprint]{acl}

\usepackage{times}
\usepackage{latexsym}

\usepackage[T1]{fontenc}

\usepackage[utf8]{inputenc}

\usepackage{microtype}

\usepackage{inconsolata}

\usepackage{graphicx}

\usepackage{amsmath}
\usepackage{booktabs}
\usepackage{multirow}

\title{Uni-Stream: A Unified Streaming Speech Language Model\\ for Speech Synthesis and Recognition}

\author{
  \textbf{Xuan Shi\textsuperscript{1}},
  \textbf{Chang Zeng\textsuperscript{2}},
  \textbf{Tiantian Feng\textsuperscript{1}},
  \textbf{Ye Zhang\textsuperscript{3}}, \\
  \textbf{Jun Liu}, 
  \textbf{Shih-Heng Wang\textsuperscript{1}},
  \textbf{Lie Lu\textsuperscript{4}},
  \textbf{Shrikanth Narayanan\textsuperscript{1}},
  \\
  \\
  \textsuperscript{1}University of Southern California, USA
  \textsuperscript{2}National Institute of Informatics, Japan \\
  \textsuperscript{3}Massachusetts Institute of Technology, USA
  \textsuperscript{4}Dolby Laboratories, USA
  \\
  \small{
    \textbf{Correspondence:} \href{xuanshi@usc.edu}{xuanshi@usc.edu}
}}
 
\begin{document}
\maketitle
\begin{abstract}
Unified streaming speech language models aim to support speech recognition and synthesis within a single model, enabling real-time human–computer interaction. However, jointly supporting these two directions of speech–text conversion while preserving task-specific performance remains challenging. We propose Uni-Stream, a unified streaming speech language model built on compact pretrained speech models. Uni-Stream processes interleaved text and speech in a streaming format and achieves performance comparable to task-specific models through two-stage training. We also develop a data annotation pipeline for streaming training and will publicly release the generated transcriptions and word-level timestamps.
We further examine Uni-Stream's representations across tasks and modalities and find that speech and text residence in distinct latent feature space even when trained in a unified model architecture, while representations within each modality become more task-dependent in later layers.

\end{abstract}

\section{Introduction}

Unified streaming speech language models integrate automatic speech recognition (ASR) and text-to-speech (TTS) synthesis within a single model, facilitating real-time human--computer interaction and low-cost model deployment applications. However, jointly training a model on both tasks may potentially degrade performance relative to task-specific models in performing ASR or TTS. A central challenge is therefore to combine recognition and synthesis in a streaming setting with low latency while retaining the capabilities of each task-specific model.

Prior work has pursued unification and streaming largely separately. Unified speech language models perform recognition and synthesis with a shared backbone \citep{yan2025minguniaudio,guan2025univoice,li2026holitok,cai2026gpa}, but most operate offline and require complete speech or text inputs. Streaming recognition and synthesis systems process their input incrementally, either by interleaving text and speech in one sequence \citep{du2024cosyvoice2,bai2025speakstream,tu2026vibevoice_asr_streaming} or by delaying one stream relative to the other \citep{defossez2024moshi,zeghidour2025dsm}, with a separate model trained for each task.
\citet{chen2021speecht} developed a unified streaming model for TTS and ASR based on a neural transducer, but their evaluation was limited to single-speaker data.
To our knowledge, no prior work has demonstrated both recognition and synthesis with streaming input using a single shared decoder-only language-model backbone.

In this work, we develop Uni-Stream, a unified streaming speech language model, through three design choices: streaming speech--text interleaving, pretrained backbone selection, and two-stage training. For speech--text interleaving, we compare fixed-ratio and time-aligned formats in TTS. Time-aligned interleaving yields lower TTS word error rates in our streaming adaptation experiments, motivating its use in the unified model. To build on pretrained capabilities for both ASR and TTS, we select the backbone from the VibeVoice family \citep{peng2025vibevoice_tts,peng2026vibevoice_asr}, whose TTS and ASR variants use similar architectures, especially for the identical speech tokenizer. We first adapt separate TTS and ASR models to our streaming formats, then initialize Uni-Stream from the adapted TTS model and jointly train it on both tasks. We formulate task-specific distillation objectives to preserve synthesis capabilities and facilitate recognition learning, and examine their contributions through training ablations. Our evaluation shows that the unified model achieves performance comparable to the task-specific models on both TTS and ASR. To support training in the time-aligned format, we develop a transcript voting and word-level time-alignment pipeline, retaining approximately 40,000 hours of Emilia-EN for training after filtering. We will release voted transcriptions for Emilia-EN and word-level timestamps for Emilia-EN, GigaSpeech, and LibriSpeech to support reproducibility and further research.

Beyond task performance, we ask how the unified model represents speech and text. Does learning both speech-to-text and text-to-speech mappings bring their internal representations closer together? To investigate this question, we probe representations across tasks, modalities, and layers. We find that speech and text representations remain distinct across layers despite joint training in both directions. Across tasks, representations are similar in early layers but diverge in later layers.

Our contributions are threefold:
\begin{itemize}
    \item We develop a unified streaming speech language model for both ASR and TTS with a set of shared weights, achieving performance comparable to task-specific streaming models on both tasks.
    \item We show through representation probing that speech and text remain distinct, while TTS and ASR representations are similar in lower layers and diverge in upper layers.
    \item We transcribe and have time-aligned segmentation for 40,000 hours of English data, facilitating the future time-aligned streaming modeling research. 
\end{itemize}

\section{Related Work}

\subsection{Speech Language Models}

Language-model-based speech recognition systems generate transcripts from encoded speech \citep{ma2024slam_asr,Qwen3-ASR,peng2026vibevoice_asr}. Language-model-based speech synthesis systems generate speech conditioned on text \citep{chen2025valle,wang2024hamtts,du2024cosyvoice2,peng2025vibevoice_tts}. Unified models perform both tasks with a single language model, operating on discrete speech tokens \citep{cai2026gpa} or on continuous representations \citep{guan2025univoice,yan2025minguniaudio,li2026holitok}. Balancing the two capabilities remains delicate: recognition and synthesis place competing demands on speech representations \citep{yan2025minguniaudio}, and starting the generation module from a synthesis-trained checkpoint improves synthesis but can reduce recognition accuracy \citep{li2026holitok}. The results reported for these models, some with progressive output, do not establish that both tasks run on streaming input.

Streaming means processing the input as it arrives, with bounded lookahead, not only emitting output progressively. Serial interleaving places text and speech in one sequence: recognition alternates audio chunks with incremental transcripts \citep{tu2026vibevoice_asr_streaming}, and synthesis alternates text and speech at a fixed ratio \citep{du2024cosyvoice2} or at word boundaries with text lookahead \citep{bai2025speakstream}; our time-aligned format follows the latter. Delayed-stream models instead keep text and audio in parallel and use separately trained recognition and synthesis models with opposite delay directions \citep{defossez2024moshi,zeghidour2025dsm}. Full-duplex dialogue models share a backbone for listening and speaking, with an emphasis on conversational interaction \citep{defossez2024moshi,zhang2025omniflatten}. An early transducer, Speech-T, was trained jointly for synthesis and recognition with streaming in both directions on a single-speaker corpus \citep{chen2021speecht}. We study how to extend a pretrained speech synthesis model to both tasks with streaming input while retaining the quality of our separately adapted streaming models.

\subsection{Speech--Text Representations in Speech Language Models}

Speech shares linguistic content with text while carrying additional information about prosody, speaker identity, and acoustic context.
Understanding how these shared and modality-specific attributes are represented is important to joint speech--text modeling.
Evidence from multimodal language models suggests that non-textual inputs can acquire language-aligned representations in intermediate or later layers \citep{wu2025semantic,neo2025towards,nikankin2026same}.
However, studies of cross-model alignment caution that semantic overlap and functional compatibility do not necessarily imply a common representational structure \citep{koepke2026back,smith2025functional}.
The relationship between cross-modal alignment and effective information use remains an open question.

In speech language models, this question spans both tokenization and language modeling.
Codec probing reveals that phonetic, lexical-semantic, and accent information are encoded unevenly across representations, highlighting the importance of tokenizer design \citep{shi2026speech,wang2026towards}.
Yet retaining information does not ensure its effective use: acoustic attributes can remain recoverable from LM hidden states even when the output head fails to express them in task predictions \citep{jo2026does}.
These findings motivate examining both what speech representations preserve and how the LM accesses that information.


\section{Uni-Stream: Unified Streaming Speech Language Model}

\subsection{Overview}

We propose Uni-Stream, a unified streaming speech language model that performs text-to-speech synthesis (TTS) and automatic speech recognition (ASR) with shared weights. To this end, we design the model from three key aspects: streaming speech–text interleaving data format, pretrained backbone selection, and two-stage training.

\begin{figure*}[t]
  \centering
  \includegraphics[width=\textwidth]{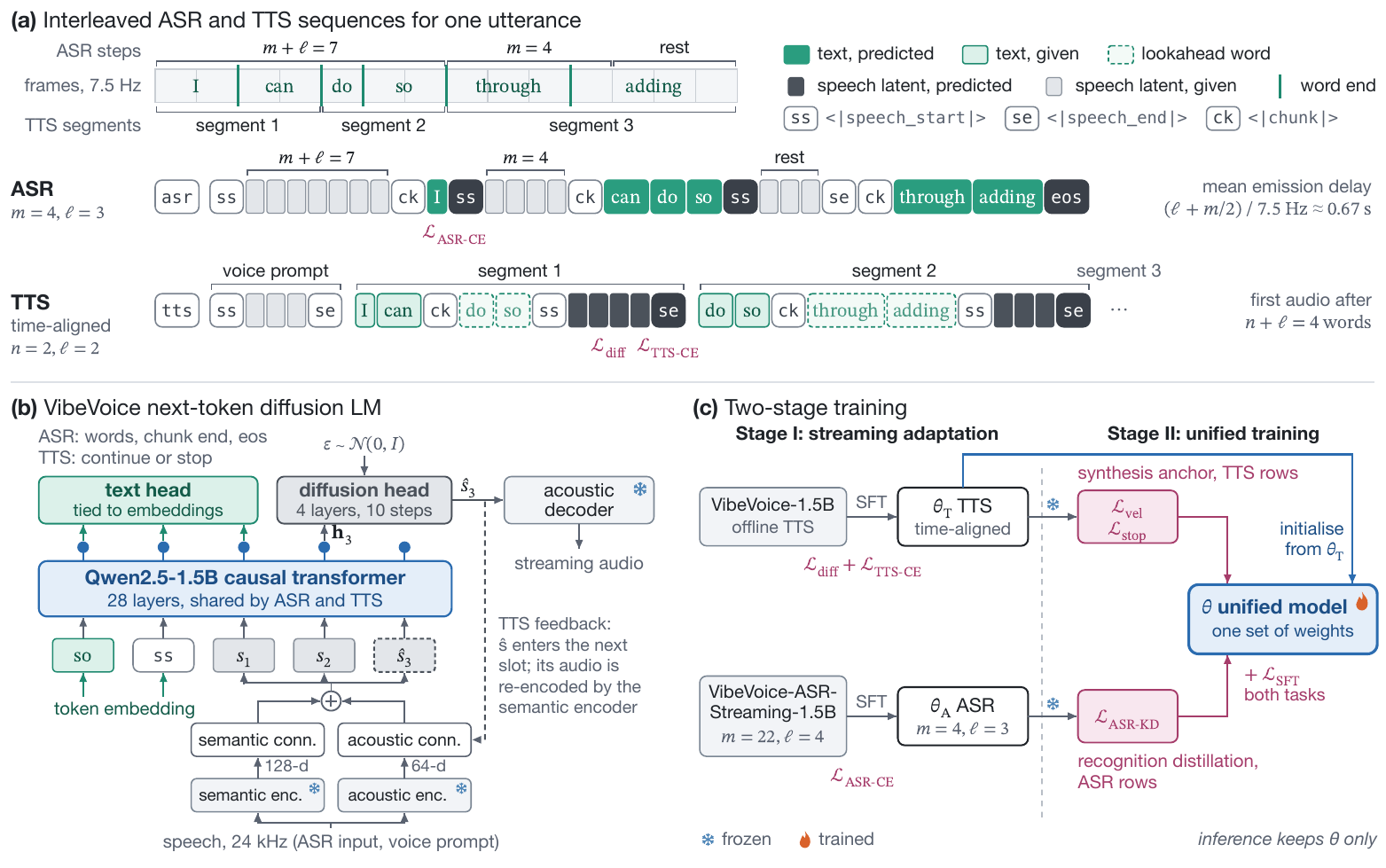}
  \caption{Overview of Uni-Stream.
  (a) Streaming ASR and time-aligned TTS sequences (frame counts illustrative).
  (b) Shared backbone with text and diffusion heads.
  (c) Two-stage training: task-specific streaming adaptation followed by joint training. The distillation branches show the objectives evaluated in our training ablations.}
  \label{fig:unified_slm}
\end{figure*}

As shown in Figure~\ref{fig:unified_slm}, each input sequence begins with a task token specifying TTS or ASR, followed by interleaved text and speech chunks. We build on the VibeVoice model family \citep{peng2025vibevoice_tts,peng2026vibevoice_asr,tu2026vibevoice_asr_streaming} for its synthesis and recognition capabilities. Our two-stage training procedure first adapts separate TTS and ASR models to our streaming formats, then initializes Uni-Stream from the adapted TTS model and jointly trains it on both tasks. We introduce synthesis anchoring and recognition distillation into unified training objectives and evaluate their contributions through ablations.

\subsection{Streaming Speech-Text Interleaving}
\label{sec:data_formatting}

Streaming ASR interleaves incoming speech chunks with incremental transcriptions. For streaming TTS, we consider fixed-ratio interleaving \cite{du2024cosyvoice2} and time-aligned interleaving \cite{bai2025speakstream}. We use $S_i$ and $T_i$ to denote speech and text chunks. Throughout the paper, $m$ denotes the speech-frame step size, $n$ the text step size (tokens or words, depending on the format), and $\ell$ the lookahead (speech frames for ASR and words for time-aligned TTS).

\paragraph{ASR}
Given speech frames, we advance by $m$ frames per step with $\ell$ frames of lookahead. Each $T_i$ contains the words whose end times fall in the $i$-th $m$-frame interval. The ASR sequence is
\begin{equation}
[S_1, T_1, S_2, T_2, \ldots].
\end{equation}
Before end-of-utterance truncation,
\begin{equation}
\begin{aligned}
S_1 &= s_{1:m+\ell},\\
S_i &= s_{(i-1)m+\ell+1:im+\ell},\quad i>1.
\end{aligned}
\end{equation}
The first step consumes $m+\ell$ frames; each subsequent step adds $m$ frames without repeating lookahead frames. When the audio ends, the model emits the remaining text.

\paragraph{TTS}
Given a text sequence, we alternate text chunks with generated speech chunks. After the voice prompt, the TTS sequence is
\begin{equation}
[T_1, S_1, T_2, S_2, \ldots].
\end{equation}

\emph{Fixed-ratio.} Each step pairs $n$ text tokens with $m$ speech frames. With $t_j$ denoting a text token, the chunks are
\begin{equation}
\begin{aligned}
T_i &= t_{(i-1)n+1:in},\\
S_i &= s_{(i-1)m+1:im}.
\end{aligned}
\end{equation}
Chunk sizes are fixed independently of word timing. Once the text reaches the end, speech generation continues until the utterance is complete.

\emph{Time-aligned.} Here, text chunks advance by $n$ words rather than tokens, with $\ell$ additional words as lookahead. With $w_j$ denoting a word, the text window is
\begin{equation}
T_i = w_{(i-1)n+1:in+\ell}.
\end{equation}
$S_i$ is the corresponding speech chunk for the first $n$ words of $T_i$. The remaining lookahead words provide context and reappear in the next window. Alignment determines where each speech chunk ends during training; the model predicts these boundaries at inference. Appendix~\ref{sec:data_construction} details alignment and boundary handling.

\subsection{Unified Streaming Backbone}

In this work, we combine representations from the TTS and ASR variants of the VibeVoice family to form the speech input to our unified model. Specifically, both VibeVoice-TTS~\citep{peng2025vibevoice_tts} and VibeVoice-ASR~\citep{peng2026vibevoice_asr} use acoustic and semantic tokenizers that encode 24\,kHz speech into 7.5 latent frames per second. These low frame-rate representations have been shown to effectively capture acoustic and semantic information.

Using these combined speech representations, VibeVoice-ASR autoregressively generates text during inference, while VibeVoice-ASR-Streaming \citep{tu2026vibevoice_asr_streaming} extends it to streaming recognition. VibeVoice-TTS uses next-token diffusion, combining an autoregressive backbone with a diffusion head \citep{peng2025vibevoice_tts}. 
During inference, the backbone encodes the input text and preceding speech into a hidden state that conditions the diffusion head to denoise the next acoustic latent. 
To support both recognition and synthesis within a single model, we adopt the VibeVoice-TTS backbone with separate text and speech output heads for ASR and synthesis, respectively.

\subsection{Two-Stage Training}

Stage~I adapts separate TTS and ASR models to the streaming formats. These specialists provide task-specific performance references and distillation targets for Stage~II, which jointly trains Uni-Stream on synthesis and recognition.

\paragraph{Stage I: Task-specific Streaming Adaptation.}

We first fine-tune VibeVoice-TTS and VibeVoice-ASR-Streaming independently to adapt them to our streaming interleaving formats while limiting deviations from their pretrained capabilities.
For TTS, we train two separate models for each interleaving format: fixed-ratio and time-aligned, using diffusion loss $\mathcal{L}_{\mathrm{diff}}$ on target speech latents and cross-entropy $\mathcal{L}_{\mathrm{TTS\text{-}CE}}$ at frame-continuation and speech-ending positions.
For ASR, we compute the cross-entropy loss $\mathcal{L}_{\mathrm{ASR\text{-}CE}}$ over transcript tokens, text-chunk terminators, and the end-of-sequence token.

\paragraph{Stage II: Unified Synthesis and Recognition.}
Let $\theta_{\mathrm{T}}$ and $\theta_{\mathrm{A}}$ denote the adapted Stage~I TTS and ASR specialists. We initialize Uni-Stream $\theta$ from $\theta_{\mathrm{T}}$ and train on both tasks with the supervised objective:
\begin{equation}
\mathcal{L}_{\mathrm{SFT}} = \mathcal{L}_{\mathrm{diff}} + \mathcal{L}_{\mathrm{TTS\text{-}CE}} + \mathcal{L}_{\mathrm{ASR\text{-}CE}}.
\label{eq:sft}
\end{equation}
Each term is averaged over its target frames or token positions within the corresponding task.
Joint training can degrade the TTS capabilities inherited from $\theta_{\mathrm{T}}$ as the model learns ASR. We therefore formulate a synthesis anchor to preserve inherited TTS capability, together with recognition distillation to facilitate ASR learning. Both specialists, $\theta_{\mathrm{T}}$ and $\theta_{\mathrm{A}}$, are frozen in the teacher forcing training.  

\paragraph{Synthesis anchor.}
\label{sec:synthesis_anchor}
The synthesis anchor works as the TTS teacher for providing acoustic predictions and stopping decisions. Teacher model $\theta_{\mathrm{T}}$ and student model $\theta$ receive the same TTS sequence, but compute their hidden states independently.

For acoustic distillation, we match diffusion velocities at the noisy latents and timesteps used for supervised training:
\begin{equation}
\begin{aligned}
\mathcal{L}_{\mathrm{vel}} &= \frac{1}{|\mathcal{F}|\,K\,d}\sum_{j\in\mathcal{F}}\sum_{k=1}^{K} \big\|\Delta_{j}^{k}\big\|_2^2,\\
\Delta_{j}^{k} &= v_\theta(\mathbf{z}_{j}^{k}, \tau_{j}^{k}, \mathbf{h}^{\theta}_{j})\\
&\quad - v_{\theta_{\mathrm{T}}}(\mathbf{z}_{j}^{k}, \tau_{j}^{k}, \mathbf{h}^{\theta_{\mathrm{T}}}_{j}),
\end{aligned}
\label{eq:anchor_vel}
\end{equation}
where $\mathcal{F}$ indexes target speech frames, $K$ is the number of noise samples per frame, and $d$ is the latent dimension. The head $v_\theta$ predicts velocity from noisy latent $\mathbf{z}_j^k$ at diffusion timestep $\tau_j^k$, conditioned on hidden state $\mathbf{h}_j^\theta$ preceding frame $s_j$.

For termination distillation, we match the two-token softmax distribution $q_\theta$ over \texttt{<|speech\_pad|>} (another frame) and \texttt{<|speech\_end|>} (end of a time-aligned chunk or a fixed-ratio utterance). Over their supervised positions $\mathcal{P}$, we minimize
\begin{equation}
\mathcal{L}_{\mathrm{stop}} = \frac{1}{|\mathcal{P}|}\sum_{p\in\mathcal{P}} \mathrm{KL}\big(q_{\theta_{\mathrm{T}}}(\cdot\mid p)\,\big\|\,q_\theta(\cdot\mid p)\big).
\label{eq:anchor_stop}
\end{equation}

\paragraph{Recognition distillation.}
To facilitate ASR learning, we define a distillation loss that matches the ASR teacher's next-token predictions, supplementing the supervised cross-entropy $\mathcal{L}_{\mathrm{ASR\text{-}CE}}$.
At each position, the teacher $\theta_{\mathrm{A}}$ and student $\theta$ receive the same speech context and reference text prefix and use identical chunk and lookahead settings. 
Let $r_\theta(\cdot\mid p)$ denote the full-vocabulary next-token distribution and $\mathcal{A}$ the positions supervised by $\mathcal{L}_{\mathrm{ASR\text{-}CE}}$. We minimize:
\begin{equation}
\mathcal{L}_{\mathrm{ASR\text{-}KD}} = \frac{1}{|\mathcal{A}|}\sum_{p\in\mathcal{A}} \mathrm{KL}\big(r_{\theta_{\mathrm{A}}}(\cdot\mid p)\,\big\|\,r_\theta(\cdot\mid p)\big).
\label{eq:asr_distill}
\end{equation}
Therefore, the Stage~II objective can be written as
\begin{equation}
\begin{aligned}
\mathcal{L} &= \mathcal{L}_{\mathrm{SFT}} + \lambda_{\mathrm{vel}}\,\mathcal{L}_{\mathrm{vel}}\\
&\quad + \lambda_{\mathrm{stop}}\,\mathcal{L}_{\mathrm{stop}} + \lambda_{\mathrm{ASR}}\,\mathcal{L}_{\mathrm{ASR\text{-}KD}},
\end{aligned}
\label{eq:anchor_total}
\end{equation}
where enabled distillation terms have unit weight. Section~\ref{sec:stage2} compares training with neither teacher, the TTS teacher alone, and both teachers. The synthesis anchor applies to TTS samples and recognition distillation to ASR samples. Only the unified model is used at inference.

\section{Experiments}

\subsection{Experimental Setup}
\label{sec:experimental_setup}

\paragraph{Data preparation.}
We use the English subset of Emilia (Emilia-EN) \citep{he2024emilia} for both TTS and ASR, supplemented with GigaSpeech XL \citep{chen2021gigaspeech} and LibriSpeech train-960 \citep{panayotov2015librispeech} for ASR.
To construct streaming training sequences, our pipeline produces transcripts with word-level timestamps.
For Emilia-EN, considering its transcription is generated by WhisperX \cite{bain2023whisperx} and not necessary to be perfect, we combine the original transcripts with two ASR hypotheses through majority voting to refine the transcription, then merge word-boundary estimates from three alignment sources.
We filter Emilia-EN utterances by transcript agreement, alignment quality, and chunk-length constraints, retaining approximately 40k hours for training with development and test speakers held out.
For GigaSpeech and LibriSpeech, we normalize the provided transcripts and obtain word timestamps with Qwen3-ForcedAligner \citep{Qwen3-ASR}.
Word-end times determine speech-segment boundaries for time-aligned TTS and text-emission chunks for streaming ASR.
Appendix~\ref{sec:appendix_data} details annotation, filtering, and sequence construction.

\paragraph{Training.}
Stage~I adapts VibeVoice-1.5B TTS and VibeVoice-ASR-Streaming-1.5B to our streaming TTS and ASR formats, respectively, on a 10k hours of speech subset from the full training set.
Stage~II jointly trains TTS and ASR for 300k steps, using the full training set with 40k hours of Emilia for TTS and a mixture of 28k hours of Emilia, 10k hours of GigaSpeech, and two passes over the 960-hour LibriSpeech training set for ASR, approximately balancing the two tasks by speech duration.
Both stages use AdamW with a peak learning rate of $10^{-5}$ and cosine decay, keeping the speech tokenizers frozen.

\paragraph{Streaming settings.}
Following Section~\ref{sec:data_formatting}, ASR uses $m=4$ speech frames per step with $\ell=3$ lookahead frames. Time-aligned TTS uses $n=2$ words per step with $\ell=2$ lookahead words, while fixed-ratio TTS pairs $n=5$ text tokens with $m=6$ speech frames, following VibeVoice's streaming TTS implementation.\footnote{\url{https://github.com/microsoft/VibeVoice/blob/main/vibevoice/modular/modeling_vibevoice_streaming_inference.py}}
TTS inference uses 10 diffusion steps with text and voice CFG scales of 3.0 and 4.0, respectively. ASR uses greedy decoding.

\paragraph{Evaluation and baselines.}
We evaluate TTS on Seed-TTS-Eval English \citep{anastassiou2024seedtts}, reporting word error rate (WER) computed with Whisper-large-v3 \citep{radford2023whisper}, speaker similarity (SIM) with WavLM \cite{chen2022wavlm}, and UTMOS \citep{saeki2022utmos}.
We evaluate ASR using corpus-level WER on LibriSpeech test-clean and test-other.
Streaming metrics include TTS first-packet latency, ASR mean word-emission delay (excluding computation), and real-time factor (RTF) for both tasks.

To our knowledge, no prior unified speech language model supports both streaming TTS and ASR, we therefore compare with offline unified models, HoliTok-Unite \citep{li2026holitok} and Ming-UniAudio \citep{yan2025minguniaudio}, and with single-task streaming models: Nemotron Speech Streaming \citep{nvidia2026nemotron}, VibeVoice-ASR-Streaming-7B \citep{tu2026vibevoice_asr_streaming}, and Kyutai DSM-ASR \citep{zeghidour2025dsm} for ASR, and Kyutai DSM-TTS \citep{zeghidour2025dsm} and VibeVoice-Realtime-0.5B \citep{microsoft2025vibevoicerealtime} for TTS.  

Training and evaluation details are provided in Appendix~\ref{sec:appendix_experiments}.

\subsection{Stage~I: Streaming Adaptation}
\label{sec:stage1}

We adapt the offline VibeVoice-1.5B TTS and the streaming VibeVoice-ASR-Streaming-1.5B models to our streaming formats in Section~\ref{sec:data_formatting}.

\begin{table}[t]
\centering
\caption{Stage~I ASR results on LibriSpeech test sets.}
\resizebox{\columnwidth}{!}{%
\begin{tabular}{lccccc}
\toprule
\multirow{2}{*}[-0.5ex]{Model} & \multicolumn{2}{c}{WER (\%)$\downarrow$} & \multicolumn{2}{c}{Emit delay (s)$\downarrow$} & \multirow{2}{*}[-0.5ex]{RTF$\downarrow$} \\
\cmidrule(lr){2-3}\cmidrule(lr){4-5}
                        & clean & other & clean & other &      \\
\midrule
VibeVoice-ASR-Streaming & 3.54  & 9.45  & 1.68  & 1.65  & 0.17 \\
Ours (Adapted ASR)      & 3.56  & 10.56 & 0.66  & 0.67  & 0.19 \\
\bottomrule
\end{tabular}}
\label{tab:stage1_asr}
\end{table}

\paragraph{ASR Adaptation}

Shown in Table~\ref{tab:stage1_asr}, for ASR model, VibeVoice-ASR-Streaming-1.5B~\cite{tu2026vibevoice_asr_streaming} orignally needs $m=22$ speech frames per step with $\ell=4$ lookahead frames, which results in a word-emission delays to $1.68$ and $1.65$ s. In our streaming adaptation, we reduce the number of speech frames per step to $m=4$ and the number of lookahead frames to $\ell=3$, which reduces the word-emission delays to $0.66$ and $0.67$ s. With a significantly shorter latency than the VibeVoice-ASR-Streaming model, the adapted streaming ASR model effectively keeps a comparable WER on test-clean at 3.56\%, while increasing WER on test-other by 1.1\%.

\begin{table}[t]
\centering
\caption{Stage~I TTS results on Seed-TTS-Eval English.}
\resizebox{\columnwidth}{!}{%
\begin{tabular}{lccccc}
\toprule
Model           & WER(\%)$\downarrow$  & SIM$\uparrow$  & UTMOS$\uparrow$     & FPL(ms)$\downarrow$     & RTF$\downarrow$     \\
\midrule
Human recording & 2.14                 & 0.734          & 3.527               & --                  & --                  \\
VibeVoice-1.5B  & 3.04                 & 0.689          & 3.971               & --                  & --                \\
\midrule
Fixed-ratio     & 7.67                 & \textbf{0.720} & 3.813               & 122                 & 0.57                \\
Time-aligned    & \textbf{3.54}        & 0.712          & 4.005               & 130                 & 0.66                \\
\bottomrule
\end{tabular}}
\label{tab:stage1_tts}
\end{table}

\paragraph{TTS Adaptation}
After adaptation on approximately 10k hours of speech, both interleaving formats improve speaker similarity over the reported VibeVoice-1.5B baseline (Table~\ref{tab:stage1_tts}).
Time-aligned TTS has substantially lower WER than fixed-ratio TTS (3.54\% vs.\ 7.67\%), with a small reduction in speaker similarity and slightly higher first-packet latency.
Therefore, we use time-aligned TTS format for the following Uni-Stream experiments.

\paragraph{Analysis of the TTS Interleaving Format}
\label{sec:analysis_format}

We inspect generated speech to understand the WER gap between the fixed-ratio and time-aligned formats.
Of the 39 fixed-ratio outputs with WER above 50\%, 29 complete the target text but continue speaking, which points to stopping behavior as a major source of errors.
The formats provide different stopping supervision: time-aligned training ends each segment after its assigned words, whereas fixed-ratio training provides an utterance-level stop signal, often long after the final text chunk.
These observations suggest that local stopping supervision contributes to the lower WER of time-aligned TTS.
Appendix~\ref{sec:appendix_format_analysis} provides additional stopping statistics and confidence intervals.

\subsection{Stage~II: Unified Streaming TTS and ASR}
\label{sec:stage2}

\begin{table*}[t]
\centering
\caption{Stage~II ASR and TTS results on LibriSpeech and Seed-TTS-en, respectively, at 40k and 300k training steps.}
\resizebox{\textwidth}{!}{%
\begin{tabular}{lcccccccccc}
\toprule
\multirow{3}{*}{Training configuration} & \multicolumn{5}{c}{ASR (LibriSpeech)} & \multicolumn{5}{c}{TTS (Seed-TTS-en)} \\
\cmidrule(lr){2-6}\cmidrule(lr){7-11}
 & \multicolumn{2}{c}{WER (\%)$\downarrow$} & \multicolumn{2}{c}{Emission delay (s)$\downarrow$} & \multirow{2}{*}{RTF$\downarrow$} & \multirow{2}{*}{WER (\%)$\downarrow$} & \multirow{2}{*}{SIM$\uparrow$} & \multirow{2}{*}{UTMOS$\uparrow$} & \multirow{2}{*}{FPL (ms)$\downarrow$} & \multirow{2}{*}{RTF$\downarrow$} \\
\cmidrule(lr){2-3}\cmidrule(lr){4-5}
 & clean & other & clean & other & & & & & & \\
\midrule
\multicolumn{11}{l}{\emph{40k steps}} \\
SFT only                      & \underline{6.53} & \underline{16.67} & 0.66 & 0.66 & 0.16 & 5.03 & 0.688 & 3.980 & 123 & 0.64 \\
+ TTS teacher                 & 6.76 & 16.88 & 0.66 & 0.66 & 0.18 & \underline{4.31} & \textbf{0.702} & \underline{3.995} & 125 & 0.71 \\
+ TTS \& ASR teachers          & \textbf{6.13} & \textbf{15.65} & 0.66 & 0.67 & 0.15 & \textbf{4.17} & \underline{0.698} & \textbf{4.014} & 120 & 0.59 \\
\midrule
\multicolumn{11}{l}{\emph{300k steps}} \\
SFT only                      & \underline{3.66} & \underline{10.71} & 0.66 & 0.66 & 0.17 & 4.35 & 0.713 & 3.891 & 120 & 0.59 \\
+ TTS teacher                 & \textbf{3.63} & 10.74 & 0.66 & 0.66 & 0.16 & \textbf{3.20} & \textbf{0.714} & \textbf{3.994} & 129 & 0.64 \\
+ TTS \& ASR teachers          & 3.68 & \textbf{10.56} & 0.66 & 0.66 & 0.17 & \underline{3.40} & 0.713 & \underline{3.988} & 120 & 0.61 \\
\bottomrule
\end{tabular}}
\label{tab:stage2}
\end{table*}

In Stage~II, we initialize Uni-Stream from the Stage~I TTS model $\theta_{\mathrm{T}}$ and jointly fine-tune it on TTS and ASR. We compare three training configurations of Uni-Stream to examine how synthesis anchoring and additional recognition distillation affect early learning and final performance.

Using the objective in Eq.~\ref{eq:anchor_total}, the three configurations set $(\lambda_{\mathrm{vel}},\lambda_{\mathrm{stop}},\lambda_{\mathrm{ASR}})$ to $(0,0,0)$ for SFT only, $(1,1,0)$ for the TTS teacher, and $(1,1,1)$ for both teachers. All three share the same initialization and training mixture.

\paragraph{Learning ASR while retaining TTS.}
Table~\ref{tab:stage2} reports results at 40k and 300k steps, demonstrating model performance at an early training point and after one epoch full-size training.

For TTS, distillation from the TTS teacher improves synthesis quality over supervised fine-tuning alone at both checkpoints. At 40k steps, adding the TTS teacher reduces TTS WER from 5.03\% to 4.31\%, increases speaker similarity from 0.688 to 0.702, and raises UTMOS from 3.980 to 3.995. Adding the ASR teacher further reduces TTS WER and improves UTMOS at this stage, with a slight decrease in speaker similarity. At 300k steps, the configuration with only the TTS teacher achieves the best TTS point estimates across all three metrics, although its speaker similarity is nearly identical to that of the other configurations. Its TTS WER is 3.20\%, compared with 4.35\% for SFT only and 3.40\% for the configuration with both teachers. These results support the effectiveness of synthesis anchoring in retaining TTS capabilities during joint training.

For ASR, the additional ASR teacher improves early recognition performance. At 40k steps, the configuration with both teachers achieves WERs of 6.13\% and 15.65\% on test-clean and test-other, respectively, outperforming both SFT only and the configuration with only the TTS teacher. At 300k steps, however, the three configurations achieve similar recognition performance, with WERs ranging from 3.63\% to 3.68\% on test-clean and from 10.56\% to 10.74\% on test-other. The ASR teacher helps the model learn recognition early in training, but this advantage largely diminishes with continued training.

These ablations show that synthesis anchoring supports TTS quality at both checkpoints, while additional ASR distillation benefits early recognition learning without a statistically significant improvement in final ASR or TTS metrics at 300k steps (Appendix~\ref{sec:appendix_stage2}). Subsequent evaluations use Uni-Stream trained with the TTS teacher alone.

\paragraph{Streaming efficiency.}
Joint training also preserves the streaming responsiveness established in Stage~I.
Uni-Stream maintains a mean ASR word-emission delay of 0.66\,s on both test sets and a median TTS first-packet latency of 129\,ms, comparable to the specialists' 0.66--0.67\,s and 130\,ms, respectively.
The ASR and TTS RTFs are 0.16 and 0.64, respectively, enabling faster-than-real-time inference under our evaluation setup.
These results show that Uni-Stream supports both tasks within a single model while maintaining streaming responsiveness comparable to that of the Stage~I specialists.

\begin{figure*}[!t]
  \centering
  \includegraphics[width=\textwidth]{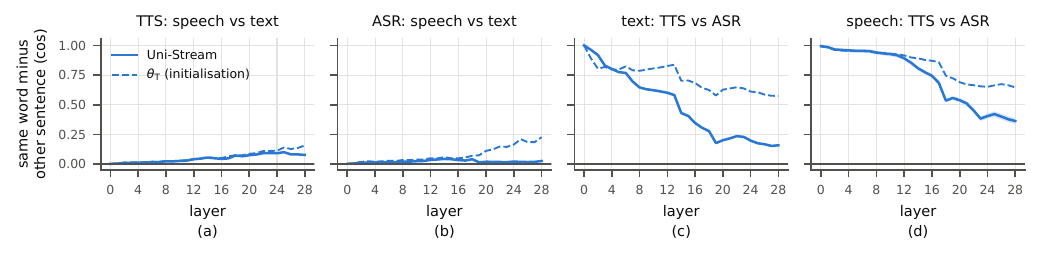}
  \vspace{-9mm}
  \caption{Content-matched similarity across layers. }
  \label{fig:analysis_sim}
  \vspace{-2mm}
\end{figure*}

\begin{table*}[h!]
\centering
\caption{Comparison with unified and single-task streaming speech models on ASR (LibriSpeech) and TTS (Seed-TTS-Eval). S: streaming (ASR latency); O: offline. n/r: not reported.}
\resizebox{\textwidth}{!}{%
\begin{tabular}{lcccccccc}
\toprule
\multirow{3}{*}{Model} & \multicolumn{2}{c}{Mode} & \multirow{3}{*}{Backbone} & \multirow{2}{*}{Data (h)} & \multicolumn{2}{c}{ASR (LibriSpeech)} & \multicolumn{2}{c}{TTS (Seed-TTS-en)} \\
\cmidrule(lr){2-3}\cmidrule(lr){6-7}\cmidrule(lr){8-9}
 & \multirow{2}{*}{ASR} & \multirow{2}{*}{TTS} & & & \multicolumn{2}{c}{WER (\%)$\downarrow$} & \multirow{2}{*}{WER (\%)$\downarrow$} & \multirow{2}{*}{SIM$\uparrow$} \\
\cmidrule(lr){6-7}
 & & & & ASR / TTS & clean & other & & \\
\midrule
\multicolumn{9}{l}{\emph{Unified models}} \\
HoliTok-Unite \citep{li2026holitok}                   & O & O & 0.5B & 60--86k / 95k & 5.48 & 12.65 & 7.20 & 0.55 \\
Ming-UniAudio \citep{yan2025minguniaudio}             & O & O & 16B-A3B & 400k / 800k & 1.62 & --    & 1.85 & 0.58 \\
\midrule
\multicolumn{9}{l}{\emph{Streaming ASR models}} \\
Nemotron Speech Streaming \citep{nvidia2026nemotron}  & S & -- & 0.6B & 530k & 2.46 & 5.07 & -- & -- \\
VibeVoice-ASR-Streaming-7B \citep{tu2026vibevoice_asr_streaming} & S & -- & 7B & $>$420k & 2.33 & 6.49 & -- & -- \\
Kyutai DSM-ASR \citep{zeghidour2025dsm}               & S & -- & 2.6B & 2.5M & 1.7 & 4.3 & -- & -- \\
\midrule
\multicolumn{9}{l}{\emph{Streaming TTS models}} \\
Kyutai DSM-TTS \citep{zeghidour2025dsm}               & -- & S & 0.3B & 88k & -- & -- & 1.58 & 0.70 \\
VibeVoice-Realtime-0.5B \citep{microsoft2025vibevoicerealtime} & -- & S & 0.5B & n/r & -- & -- & 2.05 & 0.633 \\
\midrule
Uni-Stream (ours)                                     & S & S & 1.5B & 40k / 40k & 3.63 & 10.74 & 3.20 & 0.714 \\
\bottomrule
\end{tabular}}
\\[2pt]
\label{tab:comparison}
\vspace{-2mm}
\end{table*}

\subsection{Uni-Stream Representation Analysis}
\label{sec:analysis}
\label{sec:analysis_space}

Does joint training bring speech and text representations closer in the unified model, and how does it change representations across tasks?
We probe Uni-Stream and its initialization $\theta_{\mathrm{T}}$ on ASR and time-aligned TTS sequences, investigating the cross-modality and cross-task text and speech distribution before and after unification training.
We use 64 Emilia development utterances from 26 speakers under teacher forcing.

At each layer, we average hidden states over each word's text tokens or aligned speech frames.
For each comparison, we center the two task--modality groups separately and compute same-word cosine similarity minus similarity to words from other utterances, averaging within and then across utterances.
This measures content-matched alignment after removing group offsets.
Bands show 95\% paired speaker-bootstrap confidence intervals.

Figure~\ref{fig:analysis_sim} reveals two patterns: a persistent modality gap and increasing task specificity with depth.
In panels (a, b), speech--text similarity remains low for both ASR and TTS and does not increase after joint training.
Learning both directions of speech--text conversion therefore does not bring matched speech and text representations closer under this measure.

Panels (c, d) compare TTS and ASR representations within each modality.
Cross-task similarity is high in early layers (1--9), averaging 0.78 for text and 0.96 for speech, but falls to 0.19 and 0.43 in upper layers (20--28).
To distinguish this depth-dependent pattern from changes associated with joint training, we compare Uni-Stream with its initialization at the same layers.
Early-layer scores remain close to the initialization, whereas upper-layer scores are lower by 0.42 for text and 0.24 for speech.
These results suggest that joint training preserves similar early word representations across tasks while making later representations more task-specific, even after removing average group offsets.

\subsection{Comparison among SLM Baselines}

Table~\ref{tab:comparison} compares Uni-Stream with offline unified models and task-specific streaming systems. Baseline results are taken from the corresponding papers, technical reports, or model cards.
Among the listed models, Uni-Stream uniquely supports streaming input for both recognition and synthesis with a set of shared weights.
Compared with HoliTok-Unite, Uni-Stream achieves lower WER on both tasks while supporting streaming operation, although it uses a larger backbone.
Ming-UniAudio achieves lower ASR and TTS WER, showing that unification can support stronger task performance at a larger model scale.
These comparisons provide context rather than controlled evaluations, given differences in architecture and training data.

Task-specific streaming systems still achieve better ASR performance, with the recognition gap particularly pronounced on LibriSpeech test-other.
For TTS, Uni-Stream achieves speaker similarity comparable to Kyutai DSM-TTS, but its higher WER indicates room to improve intelligibility.
The main result is therefore the combination of unified and streaming capabilities while retaining performance close to the Stage~I specialists.
The remaining gap to dedicated systems motivates further improvements in recognition robustness and synthesis accuracy.

Uni-Stream acquires streaming TTS and ASR capabilities by adapting a pretrained speech model with approximately 40k hours of training data per task.
Table~\ref{tab:comparison} describe Stage~II adaptation, not total training data, since VibeVoice's pretraining data are not reported.
The results support a practical route toward unified speech interfaces: extending pretrained capabilities to both streaming directions within one model.
Future work will focus on narrowing the quality gap to specialized systems.

\section{Conclusion}
\label{sec:conclusion}

We presented Uni-Stream, which unifies streaming speech recognition and synthesis in a shared decoder-only language model.
With time-aligned speech--text interleaving and two-stage training, Uni-Stream achieves performance comparable to the adapted task-specific models on both tasks. Our training ablations show that synthesis anchoring helps preserve TTS quality, while additional recognition distillation improves early ASR performance without a statistically significant final gain.
We will make the accompanying transcriptions and word-level timestamp annotations public to support reproducibility and further research.
Representation probing reveals a persistent modality gap: speech and text remain distinct under our similarity measure, while representations within each modality become more task-dependent in later layers.
These findings show that unified streaming capabilities can coexist with distinct modality and task representations.
Future work will focus on further improving recognition and synthesis quality.

\section*{Limitations}

While Uni-Stream presents the potential of a unified streaming speech language model for speech synthesis and recognition, several limitations and future directions remain for improving model performance, experimenting with more training configurations, and extending the model to support other languages.

\paragraph{Recognition and synthesis quality.} Although Uni-Stream performs comparably to the adapted task-specific models, the strongest specialized systems still achieve better results on LibriSpeech test-other and intelligibility in synthesized speech. Future work will investigate whether stronger pretrained backbones and different joint training strategies can improve results on these tasks while maintaining unified streaming capabilities.

\paragraph{Streaming configuration search.} Computational constraints limit our exploration of streaming and training configurations. We select chunk sizes, lookahead windows, and the text-to-speech ratio for fixed-ratio interleaving empirically or following prior work, which balances system latency and task-specific performance without an exhaustive search. 
In the future, a broader evaluation of these choices would help characterize the trade-offs between task performance, system latency, and training cost.

\paragraph{English-only evaluation.} Our data preparation, training, and evaluation are limited to English. The effectiveness of the proposed streaming formats and joint training approach in other languages remains untested. Future work will extend the data pipeline and model training to multilingual settings.

\section*{Ethics Statement}

\paragraph{Data and licensing.}
We train on publicly available speech corpora (Emilia-EN, GigaSpeech XL, and LibriSpeech train-960) and evaluate on LibriSpeech and Seed-TTS-Eval, using each corpus under its license and terms of use.
The transcriptions and timestamps described in Appendix~\ref{sec:appendix_data} are produced automatically, without human annotation.
We will publicly release these derived transcriptions and annotations under the terms of the original licenses; the release will not include audio. For Emilia, this is CC BY-NC 4.0, which restricts use to non-commercial purposes.
Emilia and GigaSpeech consist of in-the-wild recordings that contain the voices of real people and we use them only for research.

\paragraph{Potential misuse.}
Uni-Stream performs zero-shot voice cloning: given a few seconds of a speaker's voice, it can synthesize arbitrary text in that voice.
This capability could be misused for impersonation, fraud, or disinformation, and streaming synthesis makes real-time misuse, such as in phone calls, easier.
Streaming ASR could likewise facilitate unauthorized transcription of private conversations.
If we release model weights, we will restrict them to research use and require that voice prompts be used only with the speaker's consent.
We also encourage the use of synthetic speech detection and audio watermarking alongside such systems.

\paragraph{Evaluation.}
All evaluations use automatic metrics; no human subjects or crowd workers were involved.

\section*{Generative AI Disclaimer}
In this paper, AI coding agents were used to assist with development, and the authors verified all resulting code. AI writing assistants were also used to proofread the manuscript, correct grammatical errors, and improve its overall language. The authors reviewed and verified all AI-assisted text.

\bibliography{custom}

\appendix

\section{Data Preparation Details}
\label{sec:appendix_data}

\subsection{Emilia-EN Annotation}

The Emilia-EN annotations pair voted transcripts with merged word boundaries. For GigaSpeech XL and LibriSpeech train-960, timestamps correspond to the normalized provided transcripts. The annotation release described in the main text contains these derived texts and timestamps, without audio.

\paragraph{Transcript voting.}
We combine the original Emilia transcript with hypotheses from CrisperWhisper~2.0 \citep{wagner2026crisperwhisper2} and Nemotron-Streaming-ASR-0.6B \citep{noroozi2024stateful,rekesh2023fast}.
Before voting, we normalize comparison units so that numbers, abbreviations, and filler variants can match equivalent spoken forms.
We jointly align the three sequences with a beam-pruned edit alignment (beam width 32) and apply ROVER-style column-wise voting \citep{fiscus1997rover}.
Words supported by at least two hypotheses are retained; singleton insertions are removed.
Other columns retain the entry with the most support and are marked disputed; utterances are rejected if disputed columns exceed 5\% of output-producing columns.
We prefer the Emilia surface form when it matches the selected entry, and use Emilia to break ties.
Finally, we restore fillers detected only by CrisperWhisper and attach its timestamps wherever its words match the voted transcript.

\paragraph{Word-boundary merging.}
We combine CrisperWhisper timestamps with forced alignments of the voted transcript from MMS-FA \citep{pratap2024scaling} and Qwen3-ForcedAligner-0.6B \citep{Qwen3-ASR}.
Available predictions are averaged with weights proportional to inverse squared boundary-error scales.

With three predictions, an outlier is identified when the other two agree within 80\,ms and its value differs from their mean by more than 80\,ms.
Rejection is shared across each word-end/next-word-start pair: the aligner with the largest qualifying deviation on either boundary is removed from both.
Using the same surviving aligners and weights on both boundaries preserves their order when the inputs are monotone.

\paragraph{Alignment validation.}
Table~\ref{tab:timit_align} evaluates boundary accuracy on TIMIT's 5,040 utterances with manual word boundaries \citep{garofolo1993timit}.
All aligners receive the reference word sequence, and CrisperWhisper uses its forced-alignment path for this diagnostic.
For this evaluation, merge weights use the aligners' measured boundary errors. The three-way merge achieves a mean boundary error of 24.4\,ms, below that of each individual aligner.

\begin{table}[t]
\centering
\small
\caption{Word-boundary mean absolute error (ms) on TIMIT. Mean averages start and end errors.}
\begin{tabular}{lrrr}
\toprule
Aligner & Mean & Start & End \\
\midrule
CrisperWhisper & 34.9 & 31.4 & 38.5 \\
MMS-FA        & 37.1 & 43.4 & 30.8 \\
Qwen3-FA      & 30.4 & 30.0 & 30.9 \\
Three-way merge & \textbf{24.4} & \textbf{24.9} & \textbf{23.9} \\
\bottomrule
\end{tabular}
\label{tab:timit_align}
\end{table}

\paragraph{Filtering.}
We require monotone word spans, including a start no later than the end of each word.
The MMS-FA mean word score must be at least 0.30, and at most 20\% of words may have scores below 0.20; utterances without MMS scores do not pass this gate.
We also apply the sequence-length limits below and split Emilia by speaker, keeping training and evaluation speakers disjoint.

\subsection{ASR-only Corpora}
\label{sec:asr_corpus_preparation}

GigaSpeech XL and LibriSpeech train-960 use their provided transcripts without voting.
We normalize capitalization while preserving spelled-out numbers; for GigaSpeech, we also convert punctuation tags and discard segments with non-speech tags.
Qwen3-ForcedAligner-0.6B supplies one span per word, and we retain successful alignments with monotone boundaries.
We then apply the ASR chunk-length limit below.

\subsection{Word Boundaries and Sequence Construction}
\label{sec:data_construction}

For the streaming formats in Section~\ref{sec:data_formatting}, both tasks map word-end times to latent frames using
\begin{equation}
b_k = \left\lfloor \frac{\lfloor \hat{t}_{k,e} f_s \rfloor}{h} \right\rfloor,
\end{equation}
where $\hat{t}_{k,e}$ is word $k$'s end time, $f_s=24{,}000$\,Hz, and the tokenizer hop is $h=3{,}200$ samples.

\paragraph{Time-aligned TTS.}
For $W$ words and $N$ frames, we form nondecreasing cuts from $b_k$, clipped to $[0,N]$, with the first cut at zero and the last at $N$.
Each word receives the frames between consecutive cuts, so every frame is assigned once and the last word includes the remaining audio tail.
The speech for each segment consists of the frames assigned to its new words; lookahead words supply text context only.
We reject utterances with any segment exceeding 32 frames ($\approx$4.3\,s), matching the inference limit.

\paragraph{ASR.}
We clip end-frame indices to $[0,N-1]$, enforce nondecreasing order, and assign word $k$ to zero-based chunk $\lfloor b_k/m\rfloor$.
At the end of an utterance, remaining words are emitted with the available audio.
We reject utterances with more than 32 text tokens in any emission chunk.

\section{Experimental Details}
\label{sec:appendix_experiments}

\subsection{Training Configuration}

\paragraph{Optimization.}
We train the backbone, modality connectors, and output heads while keeping both speech tokenizers frozen, following the initialization and data budgets in Section~\ref{sec:experimental_setup}.
AdamW uses $(\beta_1,\beta_2)=(0.9,0.99)$, $\epsilon=10^{-8}$, weight decay 0.01, and gradient-norm clipping at 1.0.
Linear warm-up precedes cosine decay; the optimizer and schedule are reset at the start of Stage~II.
Batches group examples of similar lengths under the per-GPU token budgets in Table~\ref{tab:training_settings}, including padding.

\begin{table}[t]
\centering
\small
\setlength{\tabcolsep}{3pt}
\caption{Training settings on H100 80\,GB GPUs. TA/FR: time-aligned/fixed-ratio TTS. Uni-Stream: TTS-teacher-only configuration.}
\begin{tabular}{lcccc}
\toprule
 & \multicolumn{3}{c}{Stage~I} & Stage~II \\
\cmidrule(lr){2-4}
Setting        & TTS-TA    & TTS-FR    & ASR        & Uni-Stream \\
\midrule
Training steps  & 80k     & 80k       & 20k        & 300k    \\
GPUs           & 2         & 4         & 4          & 4          \\
Tokens per GPU & 5,600     & 2,000     & 6,400      & 4,800      \\
Warm-up        & 2\%       & 2\%       & 2\%        & 0.22\%     \\
\bottomrule
\end{tabular}
\label{tab:training_settings}
\end{table}

\paragraph{TTS conditioning.}
Each training example uses a voice prompt from another utterance by the same speaker, with requested duration sampled between 3 and 10\,s.
Text conditioning and voice prompts are independently dropped with probabilities 0.2 and 0.1.
The diffusion objective uses four noise samples per target frame.
The supervised and distillation objectives and their weights follow Sections~\ref{sec:synthesis_anchor} and~\ref{sec:stage2}.

\subsection{Evaluation Protocol}

\paragraph{Quality metrics.}
TTS evaluation uses all 1,088 Seed-TTS-Eval English utterances with voice prompts capped at 10\,s.
TTS WER is averaged over utterances after lowercasing and punctuation removal; ASR WER is pooled over each LibriSpeech test set after Whisper English normalization.
SIM and UTMOS are averaged over utterances, with UTMOS computed using the UTMOS22 strong predictor.
TTS stopping decisions are greedy, and ASR speech inputs use posterior-mean latents; other decoding settings are given in Section~\ref{sec:experimental_setup}.

\paragraph{Latency and throughput.}
TTS first-packet latency runs from the generation call to the first decoded audio packet, including prompt encoding and prefix processing, on an H100 GPU.
We report the median over 100 utterances at batch size one after warm-up.
ASR word-emission delay is the consumed audio duration at commitment minus the reference word-end time for correctly recognized words; reference boundaries come from Qwen3-ForcedAligner.
This delay excludes computation time.
RTF is averaged over utterances and divides inference wall-clock time by generated audio duration for TTS or input audio duration for ASR; ASR RTF is measured on test-clean.

\subsection{Stage~II Comparisons}
\label{sec:appendix_stage2}

\paragraph{Paired uncertainty estimates.}
We use 1,000 paired bootstrap resamples of LibriSpeech speakers and Seed-TTS-Eval voice prompts, keeping utterances paired across models.
Table~\ref{tab:stage2_paired} reports final differences and 95\% confidence intervals; ASR WER pools test-clean and test-other, and WER differences are in percentage points.
The specialist comparison uses $\theta_{\mathrm{A}}$ for ASR and $\theta_{\mathrm{T}}$ for TTS.
No statistically significant differences are detected in these final comparisons.

\begin{table}[t]
\centering
\small
\setlength{\tabcolsep}{3pt}
\caption{Paired Stage~II differences with 95\% confidence intervals. The specialist comparison uses Uni-Stream trained with the TTS teacher alone.}
\begin{tabular}{lcc}
\toprule
Metric & \shortstack{Uni-Stream $-$\\specialist} & \shortstack{Both teachers $-$\\TTS teacher only} \\
\midrule
ASR WER & +0.13 & $-$0.07 \\
 & [$-$0.03, 0.29] & [$-$0.22, 0.09] \\[2pt]
TTS WER & $-$0.33 & +0.19 \\
 & [$-$1.68, 0.63] & [$-$0.75, 1.41] \\[2pt]
SIM & +0.002 & $-$0.001 \\
 & [0.000, 0.005] & [$-$0.003, 0.002] \\[2pt]
UTMOS & $-$0.010 & $-$0.006 \\
 & [$-$0.021, 0.001] & [$-$0.019, 0.005] \\
\bottomrule
\end{tabular}
\label{tab:stage2_paired}
\end{table}

\subsection{TTS Stopping Behavior}
\label{sec:appendix_format_analysis}

To supplement the high-WER examples in Section~\ref{sec:analysis_format}, we count outputs across the full Seed-TTS-Eval set that complete the target text and then produce at least three additional words.
There are 42 such outputs for fixed-ratio TTS and four for time-aligned TTS, a difference of 3.5 percentage points with a 95\% confidence interval of [2.4, 4.8] from paired bootstrap resampling by voice prompt.

\end{document}